\documentclass[aps,pre,twocolumn,longbibliography]{revtex4-1}
\usepackage{epsfig}
\usepackage{hyperref}
\usepackage{amsfonts}
\usepackage{amsmath}
\usepackage{color}

\def\gg{{\bf g}}

\def\PB{{Poisson-Boltzmann }}

\def\D{{\bf D}}

\def\dV{{\rm d^3}{\bf r}}

\def\div{{{\rm div}\; }}

\begin{document}

\title{Structural interactions in ionic liquids linked to higher order
  Poisson-Boltzmann equations}

\author{R. Blossey} \affiliation{University of Lille 1, Unit\'e de
  Glycobiologie Structurale et Fonctionnelle, CNRS UMR8576, 59000
  Lille, France.}  \author{A. C. Maggs} \affiliation{CNRS UMR7083,
  ESPCI Paris, PSL Research University, 10 rue Vauquelin, Paris,
  75005, France.}  \author{R. Podgornik} \affiliation{Department of
  Theoretical Physics, J. Stefan Institute and Department of Physics,
  Faculty of Mathematics and Physics, University of Ljubljana, SI-1000
  Ljubljana, Slovenia}

\date{\small\today}

\begin{abstract}
  We present a derivation of generalized Poisson-Boltzmann equations
  starting {from} classical theories of binary fluid mixtures, employing
  an approach based on the Legendre transform as recently applied to
  the case of local descriptions of the fluid free energy.  Under
  specific symmetry assumptions, and in the linearized regime, the \PB
  equation reduces to a phenomenological equation introduced by
  (Bazant et al., 2011), whereby the structuring near the surface is
  determined by bulk coefficients.
\end{abstract}
\maketitle

\subsection*{Introduction}

Room-temperature ionic liquids display densely packed layers with
periodically varying charge next to charged substrates, 
described in~\cite{dosch08}, and earlier e.g.~\cite{atkin07}. This
phenomenon has been observed in numerous experiments and simulations,
see e.g.~\cite{perkin10,zhou12,smith13,cheng16,smith17}; for a review,
see \cite{varela15}.

As argued already in~\cite{dosch08} the quantitative description of
the electrostatic properties of these layers, with their observed
unusual packing of cation and anions, cannot be covered by the
standard mean-field approach based on the {lattice-gas}
Poisson-Boltzmann equation.  {What is necessary is its generalization
  either in terms of structural contributions to the free energy,
  leading to micro phase separation and charge layering
  \cite{Gavish16}, or solving the lattice Coulomb gas model beyond the
  mean-field approximation \cite{Demery12a,Demery12a}.}  Recently,
however, a phenomenological Poisson-Boltzmann equation of fourth-order
has been postulated and applied to the structure of double layers in
ionic liquids~\cite{bazant11}. The origin of the higher order term is
argued to be due to correlations between the ions, {an effect known to
  fundamentally affect the behavior of ionic fluids
  \cite{Perspective}}.  Fourth-order Poisson-Boltzmann equations have
indeed appeared in other contexts, e.g.~by invoking additional ad-hoc
parameters in order to separate out the ion correlation effects on
different length scales \cite{santangelo06}. They can also appear in
phenomenological nonlocal theories, see e.g.~\cite{paillusson10},
where they can arise from specific forms of the wave-vector dependent
dielectric function.

The arguments leading to a simple fourth-order \PB equation in the
context of ionic liquids have been, however, somewhat sketchy.  The
atomistic simulations of~\cite{zhou12}, compared to this theory,
reveal deviations which should not be present if the theory can
correctly account for the correlation effects. Despite the substantial body
of work that has emerged meanwhile on the topic, a better
understanding of the underlying physics is certainly warranted.

Recently, two of us (ACM, RP) have studied (asymmetric) steric
interactions in electrostatic double layers with Legendre transform
methods~\cite{maggs16}. The starting point of this work is the free energy
of a homogeneous binary charged mixture, from which generalized
Poisson-Boltzmann equations can be systematically derived {in a form, 
valid for any assumed equation of state of the uncharged fluid}. The resulting \PB equations, 
however, remain partial differential equations of the second order, but with different types
of local nonlinearities which cover steric effects, see also the
earlier work~\cite{borukhov97,borukhov00}. Rather surprisingly,
arbitrary fluid mixtures were shown to give rise to  \PB equations, in
which one requires only knowledge of the non-electrostatic fluid
pressure as a function of the chemical potentials of the individual
components.

In this work, we present a formal derivation of generalized
Poisson-Boltzmann equations, where the appearance of higher-order
derivative terms {in the local density} is directly linked to the spatial variation in the
concentration of the ions in the binary mixture. It arises both within
a squared-gradient approach or a more general weighted density
approximation in the free energy density. Interestingly, the equation
governing the electrostatic potential appears, to lowest relevant
order, as a fourth-order partial differential equation whose
coefficients, however, in general depend on the {\sl non-electrostatic} equation of state. As
a first application of this type of equation we compare its solutions
to those of the phenomenological equation employed by Bazant et
al.~\cite{bazant11}. We show that even with these generalisations the
only input needed by the theory is again the pressure function of the
non-electrostatic problem. The phenomenological coefficients
of~\cite{bazant11} are found in terms of {this non-electrostatic} pressure function.

\subsection*{Theory}

The starting point of our discussion is the free energy expression for
a charged binary mixture, composed of two contributions,
$ {\cal F} = {\cal F}_{electro} + {\cal F}_{fluid} $, where the
{electrostatic} contribution is given by the expression
\begin{equation} \label{eq:df} {\cal F}_{electro}[\{ c_i \}, {\bf D} ]
  = \int \dV \left [ \frac{{\bf D}^2}{2\varepsilon} - \phi (\mbox{div}
    {\bf D} - e z_i c_i) \right ]
\end{equation}
while the fluid term can be either the sum of a local free energy plus
a squared gradient
\begin{equation} \label{eq:gradf} {\cal F}_{fluid}[\{c_i\}] =
  \frac{1}{2} \int \dV \left [f(\{c_i\}) - \mu_i c_i + \kappa_{ij}
    \nabla c_i \nabla c_j \right ]
\end{equation}
{where the interaction strengths $\kappa_{ij}$ are proportional to the 
  bulk correlation lengths}, or can be written in a weighted density form
\begin{equation}
  {\cal F}_{fluid}[\{c_i\}] = \int \dV\, f(w(c_i)) \label{eq:weighted}
\end{equation}
where $w$ is supposed to be a linear, nonlocal function 
\cite{curtin}. Note that, in the expressions, we have employed Einstein's
summation convention and that in what follows, we will be rather
careless in the exact symbol (and name) used for the various free
energies $\mathcal F$. We perform multiple and sometimes non-standard
Legendre transforms, while the stationary points of all the objects
that we construct have the same value.

The local contribution to eq.xs~(\ref{eq:gradf}) is the free energy of
an isothermal binary mixture with the concentrations $c_i$ and was the
basis of the discussion that in~\cite{maggs16} led to a general theory
of asymmetric steric interactions in electrostatic double layers; the
$\mu_i$ are the chemical potentials; ${\bf D}$ is the dielectric
displacement field, $\varepsilon = \epsilon\epsilon_0$ the relative
dielectric permeability, the $z_i$ the valencies of the charged
species and $\phi$ is the Lagrange multiplier which takes care of
Gauss' law, i.e.\ the electrostatic potential.

In~\cite{maggs16}, the theory on the mean-field level was transformed
into a free energy expression for the electrostatic potential
$\phi({\bf r})$. This was achieved by making use of the Legendre
transform approach developed in~\cite{ralfe,maggsxx}. The same
approach can be applied here, where in addition we further first
introduce a vector-valued variable for the concentration gradient
given by ${\bf v}_i \equiv \nabla c_i\, $ with an associated
multiplier ${\bf g}_i$. An integration by parts and regrouping of
terms in eq.~(\ref{eq:gradf}) leads to the expression
\begin{align} \label{eq:result}
  {\cal F}[\{c_i\},& {\bf D},{\bf v}_i]  =   \\
                   & \int \dV  \left(f(\{ c_i \}) - (\mu_i + ez_i\phi + \nabla \cdot {\bf g}_i)c_i \right) \nonumber \\
  + & \int \dV \left(\frac{{\bf D}^2}{2\varepsilon} + \nabla \phi
      \cdot {\bf D} + \frac{\kappa_{ij}}{2}{\bf v}_i{\bf v}_j - {\bf
      g}_i \cdot {\bf v}_i\right) \nonumber
\end{align}
We now recognise that the stationary point of this functional with
respect to the concentration leads to the Legendre transform of the
fluid free energy, thus
\begin{align} \label{eq:grad}
  {\cal F}[\phi, \{{\bf g}_i\}]  &= - \int d^3{\bf r} \, \kappa_{ij}^{-1}\frac{{\bf g}_i{\bf g}_j}{2} \\
  - & \int d^3{\bf r} (\frac{\varepsilon}{2}{(\nabla \phi)}^2 +
      P(\{\mu_i - \phi e z_i + \nabla \cdot {\bf g}_i\}) ) \nonumber
\end{align}
where $P$ is the fluid pressure expressed as a function of the
set of chemical potentials ${\mu_i}$, as discussed in~\cite{maggs16}.  It is important
to note that the Lagrange multiplier $ {\bf g}_i$ enters as an
argument to the pressure function which is in general a highly
nonlinear function.  We recall at this point the fundamental
Gibbs-Duhem relation linking derivatives of $P$ to the concentration,
\begin{equation}
  c_i = \frac{\partial P}{\partial \mu_i}.
\end{equation}

The variational equations found by taking derivatives of the free
energy eq.~(\ref{eq:grad}) have a simple first integral in
one-dimensional geometries. This integral can be found using the
standard construction of a Hamiltonian from a Lagrangian, taking the
coordinate $x$ as equivalent to the time in a particle system, while the conserved quantity is not an energy, 
but rather the total pressure in the fluid (where external charges are absent) which can be derived as
\begin{equation} \label{osmo} p = - \epsilon \frac{(\partial_x
    \phi)^2}{2} +P + \frac{ g_i g_j}{2 \kappa_{ij}} - (\partial_x g_i)
  c_i
\end{equation}
Far from any external sources, where the potential and $g_i$ become
constant, this reduces simply to the neutral fluid pressure function
$P$.

We now take a fluid free energy based on the weighted functional form
eq.~(\ref{eq:weighted}). {The weighted density approximation is a
  more sophisticated approximation than the square gradient
  approximation. It supposes that one can evaluate the free energy of
  a non-uniform fluid from the equation of state of a uniform
  system. However the argument of the uniform state function is a non-local
  average of the density in some local neighbourhood. Typically such
  methods give more robust descriptions of fluid behaviour, above all
  in the presence of strong gradients. In the most sophisticated
  theories the weighting kernel is also calculated from the
  correlations of the equilibrium fluid but here we will take a very
  simplified form of the kernel as a Yukawa form, with range
  $k_i^{-1}$. This specific choice allows us to make considerable
  simplifications in the theory}.  The total free energy of the system is then
\begin{equation}
  {\mathcal F} = \int \dV \, \left [ - \mu_i c_i + f({w_i\ast c_i}) +\frac{\D^2}{2 \varepsilon} - \phi (\div \D -
    e z_i c_i )  \right ] \nonumber
\end{equation}
Where by $w_i \ast c_i$ we mean the convolution of the concentration
field $c_i$ with the Yakawa kernel of range $k_i^{-1}$. We pull out the
weighting function as an argument of $f$ by introducing a local
density $d_i= w_i \ast c_i$:
\begin{align}
  {\mathcal F} = \int \dV  [ - \mu_i c_i + f(d_i) +& \frac{\D^2}{2 \varepsilon} - \phi (\div \D -
                                                     e z_i c_i ) \\ -& \lambda_i (
                                                                       d_i - w_i\ast c_i) ] \nonumber 
\end{align}
where again $\lambda_i$ is a Lagrange multiplier. All the
concentrations $c_i$ occur linearly allowing us to perform the
Legendre transform:
\begin{equation}
  {\mathcal F} =  - \int \dV\, \left [ \varepsilon \frac {{(\nabla \phi)}^2} {2}
    -P(\{w_i^{-1} (\mu - e z_i \phi) ) \}   \right ]
\end{equation}
It is here that the utility of the choice of a Yukawa function becomes
clear as the inverse operator is then $w^{-1} = 1 -
\nabla^2/k_i^2$.
This gives the final \PB functional for the weighted density
approximation that involves purely the electrostatic potential,
without the introduction of any supplementary degrees of freedom:
\begin{equation}
  {\mathcal F} =  \int \dV \left [ - \varepsilon \frac {{(\nabla
        \phi)}^2} {2}  -P\left ( \left  \{ (1
      -\frac{\nabla^2}{k_i^2} ) (\mu_i - e z_i \phi ) \right \} \right )  \right ]. 
  \label{eq:G2}  
\end{equation}
{Within this local density approximation} we have therefore derived a
generalisation of the \PB equation with a Laplacian within
 the pressure function, $P$. We note that eq.~(\ref{eq:G2}) leads,
on expansion in gradients, to a free energy which is a series in
$(\nabla^2\phi)$. Retaining terms up to second order leads directly to
the free energies of the form proposed in~\cite{bazant11}.


\subsection*{Lowest order elimination of the gradient multiplier}

In eq.~(\ref{eq:result}) we can expand the pressure function to first order in
$(\div \gg_i)$ to find:
\begin{equation}
F \approx \int \dV  \left ( -P - c_i \div \gg_i   - \epsilon \frac{(\nabla
  \phi)^2}{2}  - \frac { \gg_i \gg_j }{2 \kappa_{ij} } \right )    \label{eq:approx}
\end{equation}
We  now integrate by parts and solve for $\gg_i$
\begin{equation}
\gg_i \approx \kappa_{ij} \nabla c_j
\end{equation} 
Substituting back and using Gibbs-Duhem gives an effective action in
terms of only a potential variable:
\begin{align}
{\mathcal F} =&\int \dV  \left [  -P + (\nabla P_i) \frac{\kappa_{ij} }{2}(\nabla P_j)   - \epsilon \frac{(\nabla
  \phi)^2}{2} \right ] \nonumber \\
=&\int \dV  \left [ -P - \frac{(\nabla \phi)^2 }{2}  \left (\varepsilon  - \kappa_{ij}
   e^2 z_k z_l \frac{d  c_k}{d \mu_i} \frac{d c_l}   {d \mu_j} \right )
   \right ]
\label{eq:eff}
\end{align}
since $c_l$ is clearly itself a derivative of the pressure.
The lowest order correction to the Poisson-Boltzmann function has the
appearance of an effective shift in the dielectric properties of the fluid {or indeed implies that the molecular structure 
of the system renormalizes its dielectric response}.

\subsection* {Application: The symmetric electrolyte and the
 {square} gradient approximation in one dimension}

We now develop further the application of eq.~(\ref{eq:grad}), which unlike
eq.~(\ref{eq:G2}) is not directly of the \PB form, involving only a
functional of the electrostatic potential. We work with a symmetric
binary mixture and impose a one-dimensional geometry, having in mind
one and two-plate problems. {The electroneutrality condition is trivially 
satisfied since no external charge is considered in the model. }

In this case the free energy expression, eq~(\ref{eq:grad}) reduces to
\begin{align} {\cal F}[&\phi, \{\partial_x g_i\}] = \\
  - &\int dx  \left[\frac{\varepsilon}{2}{(\partial_x \phi)}^2 +
    \frac{1}{2} \overline{g}\cdot\widehat{\kappa}\cdot\overline{g}^T +
    P(\{\mu_i - ez_i \phi + \partial_x g_i\}) \right] \nonumber 
\end{align}
with $i=(+,-)$, where, explicitly, we have for the inverse matrix
$\kappa_{ij}^{-1} \equiv \widehat{\kappa}_{ij}$
\begin{eqnarray}
  \overline{g}\cdot  \widehat{\kappa} \cdot \overline{g}^T = 
  \left(g_+\,\,\, g_-\right)
  \left(
  \begin{array}{cc}
    \widehat{\kappa}_{++}  & - \widehat{\kappa}_{+-}   \\
    - \widehat{\kappa}_{+-}  & \widehat{\kappa}_{--}     
  \end{array}
                               \right)
                               \left(\begin{array}{c}
                                       g_+  \\
                                       g_-      
                                     \end{array}
  \right)\,.
\end{eqnarray}
The standard expression for the pressure comes from a lattice gas model:
\begin{equation}
  P = \frac{1}{a^3}\ln\left(1 + \frac{\gamma}{2(1-\gamma)}\left(e^{\beta m_+(x)} + e^{\beta  m_-(x)}\right) \right)\, 
\end{equation}
with
\begin{equation}
  m_{\pm}(x) = \pm ez \phi(x) + \partial_x g_{\pm}(x)\, ,
\end{equation}
and, following ref. \cite{maggs16}, the conditions on the chemical potentials are 
$ e^{\beta \mu_+} = e^{\beta \mu_-} = \frac{1}{2}[\frac{\gamma}{1 -
  \gamma}] $.  We can now vary the functional ${\cal F}[\phi, \{\partial_x g_i\}]$ with respect
to the fields $\phi, g_+$ and $g_-$. The variation with respect to
$\phi$ yields
\begin{equation}
  - {\varepsilon}\partial_x^2\phi(x) + \frac{\gamma ze}{a^3}\left[\frac{e^{\beta m_+(x)} - e^{\beta m_-(x)}}{1 - \gamma + \gamma(e^{\beta m_+(x)} + e^{\beta m_-(x)})} \right] = 0\, ,
\end{equation}
while for the two components of $\overline{g}$ one has
\begin{equation}
  \widehat{\kappa}_{++}g_+ - \widehat{\kappa}_{+-}g_- = \frac{\gamma}{a^3}\partial_x 
  \left[\frac{e^{\beta m_+(x)}}{1 - \gamma + \gamma(e^{\beta m_+(x)} + e^{\beta m_-(x)})} \right]
\end{equation}
and
\begin{equation}
  \widehat{\kappa}_{- -}g_{- -} \widehat{\kappa}_{+-}g_+- = \frac{\gamma}{a^3}\partial_x 
  \left[\frac{e^{\beta m_-(x)}}{1 - \gamma + \gamma(e^{\beta m_+(x)} + e^{\beta m_-(x)})} \right]\,. 
\end{equation}
\\

{\it Linearisation.}
It is instructive to consider the linearized equations --- we will see
that they contain already the essential physics of this theory. We
find
\begin{equation} \label{l1} - \varepsilon\partial_x^2\phi(x) +
  \frac{\gamma ze \beta}{a^3}(2ez \phi(x) + {\partial_x}g_+(x) -
  {\partial_x}g_-(x)) = 0\, ,
\end{equation}
while for the two components of $\overline{g}$ one has
\begin{equation} \label{l2} \widehat{\kappa}_{++}g_+(x) -
  \widehat{\kappa}_{+-}g_-(x) = \frac{\gamma \beta}{a^3}(ez \partial_x
  \phi(x) + \partial_x^2 g_+(x))
\end{equation}
and
\begin{equation} \label{l3} \widehat{\kappa}_{- -}g_-(x) -
  \widehat{\kappa}_{+-}g_+(x) = \frac{\gamma \beta}{a^3}(-ez\partial_x
  \phi(x) + \partial_x^2g_-(x))\, .
\end{equation}
\\
Figure~1 shows the resulting solutions to these equations in the case
of a one-plate system with constant surface potential, $\phi(0) = V$, for
the parameters indicated in the caption. In Figure 1 A, $\kappa_{++} = \kappa_{- -}$,
while the contrast between the two couplings in increased by a factor of ten in Figure
1 B. The presence of $g_+$ and $g_-$ leads to oscillations in the profile of
the electrostatic potential whose number, amplitude and extension into the bulk depend
on the contrast between the coupling parameters. 
The equation for the electrostatic potential $\phi$ looks similar to that if a 
mechanical spring (in which the position $x$ plays the role of time), with the damping term
given by the functions $g_+(x)$ and $g_-(x)$, which in their turn essentially obey coupled
diffusion equations, with the electric field acting as a source or a 
sink term. It is therefore not surprising that, as a function of the
coupling parameters $\widehat{\kappa}_{ij}$, the solution exhibits 
damped oscillations.
\begin{figure}
  \begin{center}
    \includegraphics[scale=0.40]{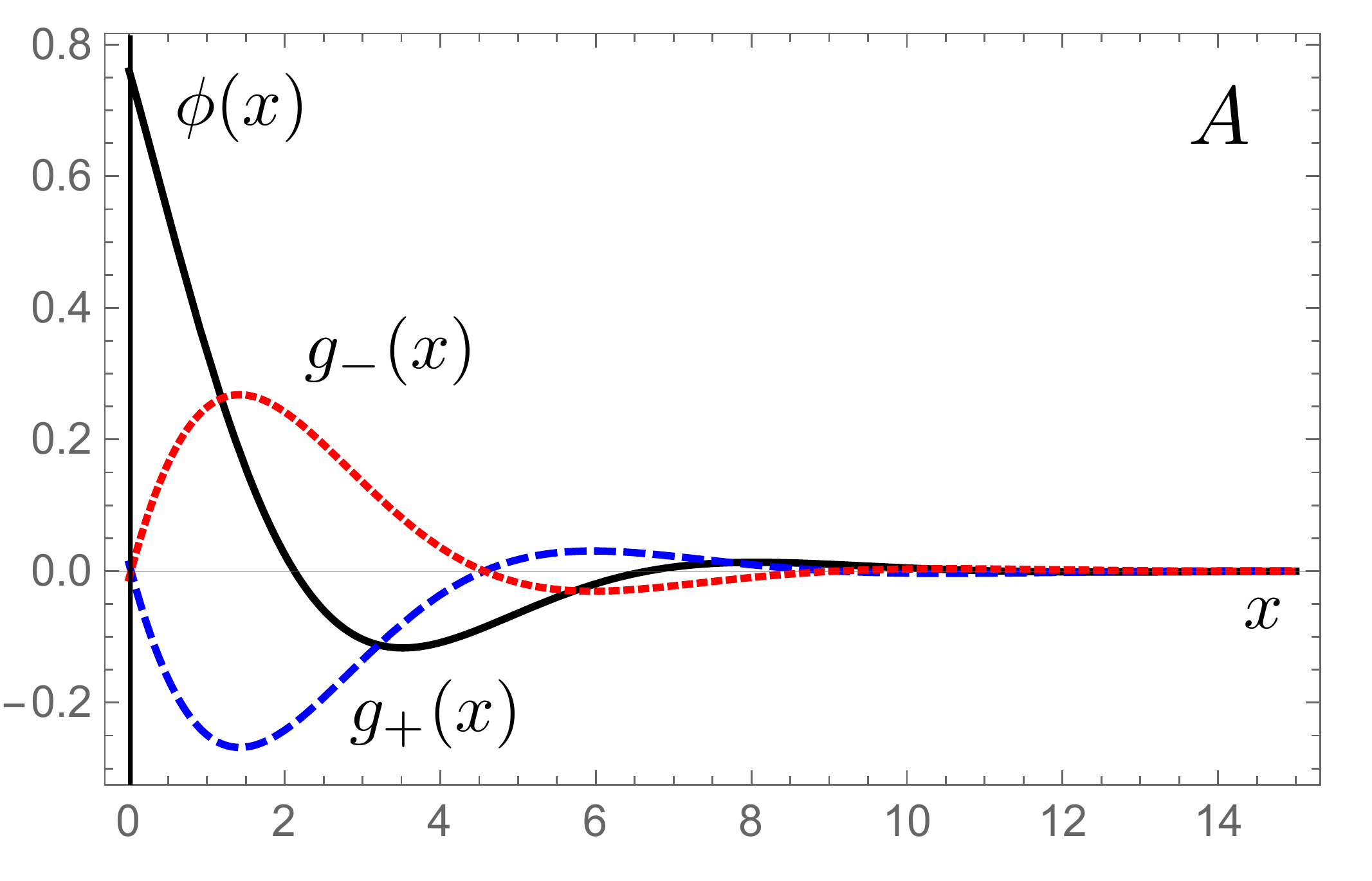}
    \includegraphics[scale=0.38]{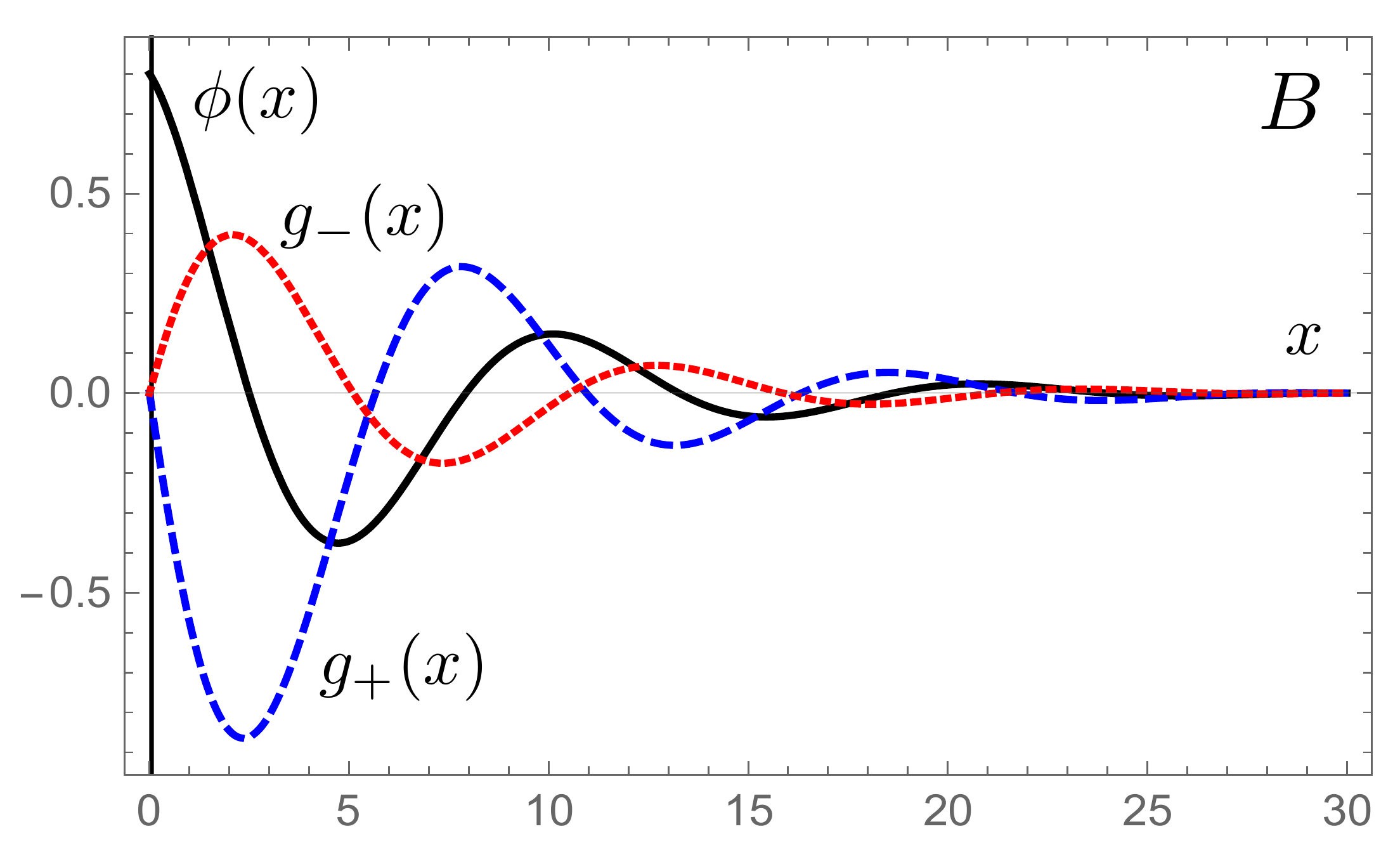}
    \caption{Oscillatory profiles of the functions $\phi$ (black) $g_+$ (blue, large dashes) and $g_-$
    (red, small dashes), obtained from
      Eqs. (\ref{l1}), (\ref{l2}), (\ref{l3}) for two different sets of parameters. The parameters in A 
      are $V=0.8$, $\kappa_{++} = \kappa_{- -} = 0.5$, $\kappa_{+ -} = 0 $. In B:  
      $V=0.8$, $\kappa_{++} = 0.05$, $\kappa_{- -} = 0.5$, $\kappa_{+ -} = 0$.
      All other parameters are set to one. The other boundary conditions are $g_+(0)=g_-(0)=0$,
      $\phi(L) = g_+(L) = g_-(L) = 0$.} 
  \end{center}
  \label{Figure1}
\end{figure}

The variational equations can be reduced to only two fields if one assumes the
symmetry condition $\widehat{\kappa}_{++} = \widehat{\kappa}_{- -} \equiv \kappa$.
We further set $\widehat{\kappa}_{+-} = 0$. 
Subtracting the equations in $g_{\pm}$ and introducing
$g \equiv g_+ - g_-$ then yields a coupled system in $\phi(x), g(x) $.
Being linear equations, one can eliminate the field $g$ in favor of $\phi$
and ends up with a forth-order equation in the electrostatic field
\begin{equation} \label{korny} \tilde{\kappa}^2\partial_x^4\psi(x)
  - \partial_x^2 \psi(x) + \psi(x) = 0\, ,
\end{equation}
where $\psi \equiv (2ez)\phi $, $ \tilde{\kappa}^2 \equiv (2(ze)^2 (\gamma \beta)^2/(\kappa \varepsilon a^6)$ 
and the spatial coordinate $x$ has been rescaled by a factor $\sqrt{\epsilon}$
with $\epsilon = \varepsilon a^3/(\gamma \beta)$.

Eq.~(\ref{korny}) is exactly the linear equation discussed in~\cite{bazant11},
identifying the coupling parameter $\tilde{\kappa}^2$ with the parameter $\delta_c^2$ used in~\cite{bazant11}.  
This connection also allows us to relate the parameter $\delta_c^2$ introduced in~\cite{bazant11} with the
interaction strength $\kappa$ in the binary mixture.  Furthermore, the linear solution derived in the 
Supplementary Material of~\cite{bazant11} can immediately be adopted by just making the
replacement $\delta_c^2$ with $\tilde{\kappa}^2$.

{\it The nonlinear case.}
In the nonlinear case, one has to solve the coupled eqs. (\ref{l1}) - (\ref{l3}). A simplification can again be made if
we consider the anti-symmetric subspace $g_+(x) = - g_-(x) \equiv \overline{g}(x) $, corresponding to antagonistic gradients in 
concentrations. In this case one has $m_+(x) = - m_-(x) \equiv m(x) $ and the resulting two equations 
for $\phi$ and $g$ can be written as 
\begin{equation} \label{nl1}
-\varepsilon \partial_x^2 \phi(x) + \frac{\gamma z e}{a^3}\frac{\sinh(\beta m(x))}{1 + 2\gamma \sinh^2(\beta m(x)/2)} = 0
\end{equation}
and
\begin{equation}   \label{nl2}
\overline{g}(x) = \frac{\gamma}{2a^3\kappa} \partial_x\left(\frac{\sinh(\beta m(x))}{1 + 2\gamma \sinh^2(\beta m(x)/2)}\right)\, . 
\end{equation}
The full nonlinear equations can easily be solved numerically as a
boundary value problem with the conditions $\phi(0)=V$, $\phi(L)=0$,
$\overline{g}(0) = \overline{g}(L) = 0$. Figure 2 shows the results for the charge density
$ P'(\phi) \equiv (1/\gamma)\tanh(\gamma \phi(x))$, for a value of
$\gamma = 0.8$ and a value of $ \phi(x=0) = V = 5$. The results
compare favorably with those obtained in~\cite{bazant11}, and therefore indicate
that our explicit assumption of antagonistic concentration gradient is implicitly present
in the theory developed there.

{\it The two-plate problem.}
Finally one can also study the two-plate problem for this equation in the similar vein as \cite{Gavish16}. 
Figure 3 plots an exemplary osmotic pressure, given by eq.(\ref{osmo}), between the plates for identical constant 
potentials at the plates, $\phi(0) = \phi(L) $, based on the linearized two-field theory discussed just before.  
Also shown is a comparison with the osmotic pressure of the linearized standard Poisson-Boltzmann equation,
which yields a parabolic potential $\phi(x) > 0$ with an exponentially 
    decaying osmotic pressure for large plate distances. In the structured fluid, upon variation of the two-plate distance $L$, the potential 
    $\phi$ first becomes flat at the center and crosses to negative values (as shown in an insert for $L=10$), resulting in a pronounced 
    minimum in $p(L)$ (not shown). 
    Upon further increase in $L$ the osmotic pressure progresses through a maximum and a second, very shallow minimum appears (see
    top insert). In this region, the electrostatic potential has a double-well structure next to the plates, as shown in the insert
    for $L=15$.

\begin{figure}
  \begin{center}
    \includegraphics[scale=0.36]{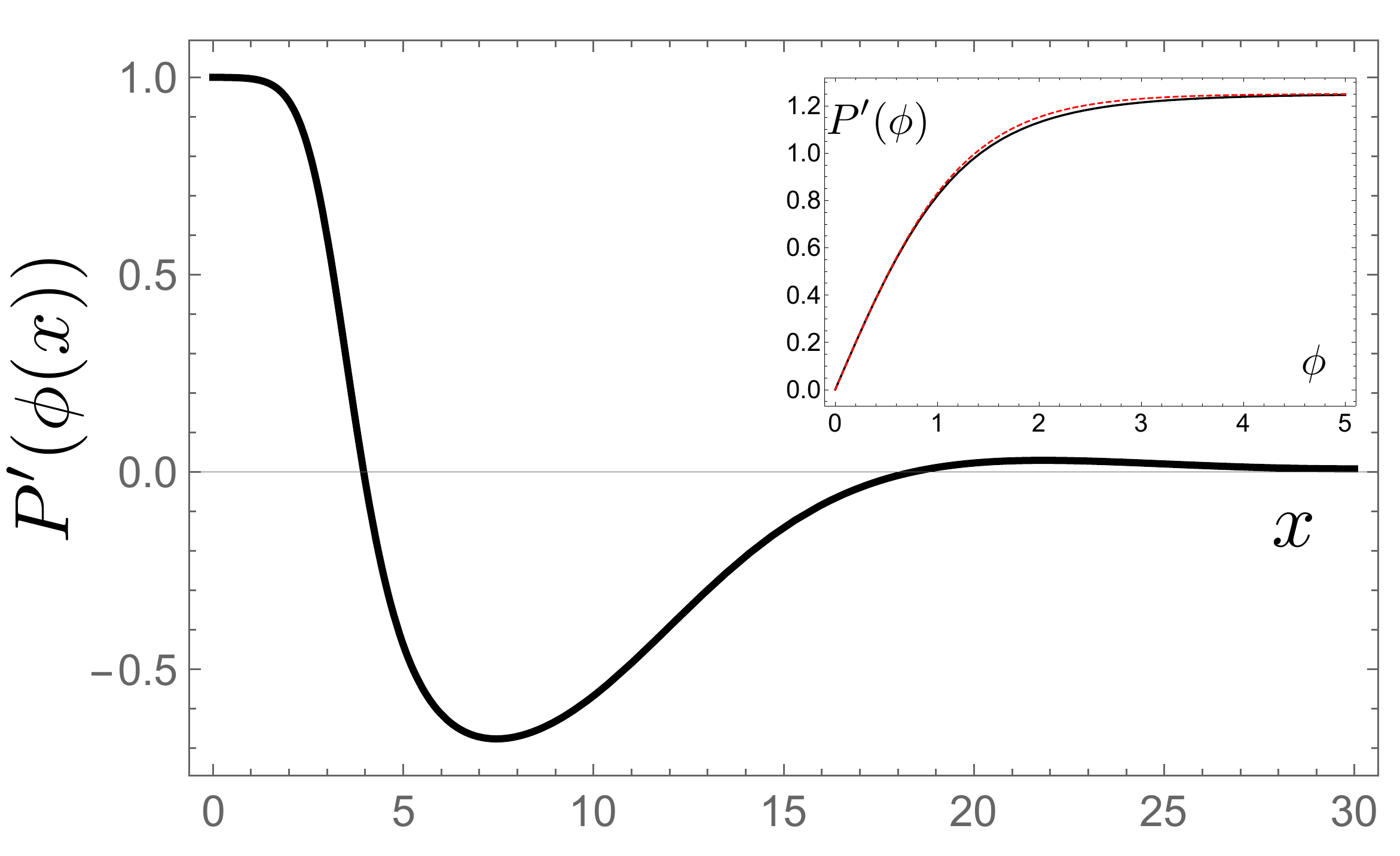}
    \caption{The charge density $P'(\phi(x))$ from the solution of $\phi(x))$obtained from eqs.(\ref{nl1}), (\ref{nl2}) for the case of constant
    potential $\phi(0) = 5$. Insert: the approximate expression used, compared to the exact expression of the pressure of the symmetric
    electrolyte (exact: black, approximation: red dashed).} 
  \end{center}
  \label{Figure2}
\end{figure}

\subsection*{Discussion}

In this work we have shown that the inclusion of concentration
gradients in a charged binary mixture leads in a natural way to the
appearance of higher-order terms in the generalized Poisson-Boltzmann equation,
which in general turns out to be much more complex than the
phenomenological fourth-order PB equation employed in \cite{bazant11}
due to the coupling of nonlinearities and spatial gradients.

{In the model case of a symmetric binary mixture the presence of
different coupling constants of the spatial gradient terms lead to pronounced 
oscillatory behaviour of the electrostatic profiles. Also, on the level of
the generalized Poisson-Boltzmann equation, the oscillations near the wall
therefore are the consequence of bulk behaviour and not of the surface coupling. 
Nonetheless, the solution of the corresponding system of
coupled nonlinear differential equations for the electrostatic
potential and the spatial gradient function yield results that are
qualitatively similar to the behaviour found in~\cite{bazant11}, and
actually coincide for the linear case, providing a physically more sound
justification of the phenomenological approach. At the same time, our
results make clear that the present state of theory does not go beyond
the mere mean-field level, and a consistent discussion of correlation
effects in ionic liquids, in addition to the structure effects, is still missing.
A description of the oscillations in electrostatic potential and osmotic pressure in 
terms of two decay lengths, as discussed in~\cite{smith17}, might 
therefore be only a special case of a more general situation in which
structural, or packing, effects play a crucial role.} 

\begin{figure}
\begin{center}
    \vspace{0.5cm}
    \includegraphics[scale=0.34]{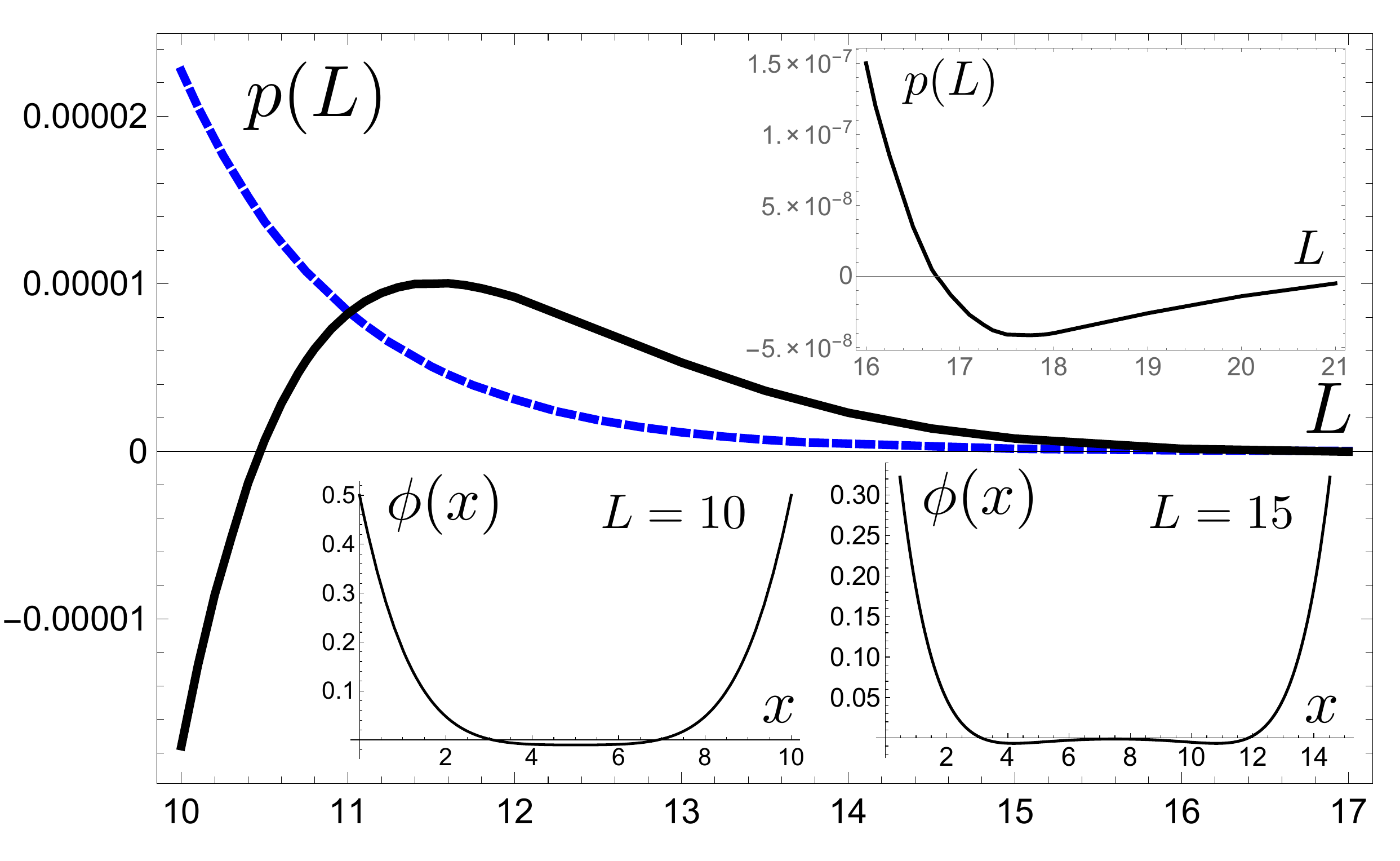}
        \caption{Osmotic pressure in the linear two-field theory (orange) and the linearized   standard Poisson-Boltzmann equation (blue), both for constant potential $\phi(x=0)=0.5$ at the plates. All parameters are set to one.  In the range of plate distances $L$ shown, the osmotic pressure shows oscillations between repulsive and attractive behaviors. The electrostatic potential, shown in two inserts for the values of $L=10$ and $L=15$, evolve from from having single minimum at mid-plate to displaying two minima next to the plates.}
 \end{center}
  \label{Figure3}
\end{figure}

{Nevertheless, it is worthwhile pointing out that the squared density gradient theory, stemming
from the non-electrostatic structural contribution to the free energy, leads at least in the linear
regime formally to the same generalized \PB equation as one would obtain from the contribution
of the non-mean-field ion correlations, approximated via an appropriate length scale 
separation \cite{santangelo06}. It would therefore seem difficult to disentangle these two effects
without a full theory taking into account the ion correlations as well as the molecular structure of the system.}


\bibliography{var}

\begin{thebibliography}{20}%
\makeatletter
\providecommand \@ifxundefined [1]{%
 \@ifx{#1\undefined}
}%
\providecommand \@ifnum [1]{%
 \ifnum #1\expandafter \@firstoftwo
 \else \expandafter \@secondoftwo
 \fi
}%
\providecommand \@ifx [1]{%
 \ifx #1\expandafter \@firstoftwo
 \else \expandafter \@secondoftwo
 \fi
}%
\providecommand \natexlab [1]{#1}%
\providecommand \enquote  [1]{``#1''}%
\providecommand \bibnamefont  [1]{#1}%
\providecommand \bibfnamefont [1]{#1}%
\providecommand \citenamefont [1]{#1}%
\providecommand \href@noop [0]{\@secondoftwo}%
\providecommand \href [0]{\begingroup \@sanitize@url \@href}%
\providecommand \@href[1]{\@@startlink{#1}\@@href}%
\providecommand \@@href[1]{\endgroup#1\@@endlink}%
\providecommand \@sanitize@url [0]{\catcode `\\12\catcode `\$12\catcode
  `\&12\catcode `\#12\catcode `\^12\catcode `\_12\catcode `\%12\relax}%
\providecommand \@@startlink[1]{}%
\providecommand \@@endlink[0]{}%
\providecommand \url  [0]{\begingroup\@sanitize@url \@url }%
\providecommand \@url [1]{\endgroup\@href {#1}{\urlprefix }}%
\providecommand \urlprefix  [0]{URL }%
\providecommand \Eprint [0]{\href }%
\providecommand \doibase [0]{http://dx.doi.org/}%
\providecommand \selectlanguage [0]{\@gobble}%
\providecommand \bibinfo  [0]{\@secondoftwo}%
\providecommand \bibfield  [0]{\@secondoftwo}%
\providecommand \translation [1]{[#1]}%
\providecommand \BibitemOpen [0]{}%
\providecommand \bibitemStop [0]{}%
\providecommand \bibitemNoStop [0]{.\EOS\space}%
\providecommand \EOS [0]{\spacefactor3000\relax}%
\providecommand \BibitemShut  [1]{\csname bibitem#1\endcsname}%
\let\auto@bib@innerbib\@empty
\bibitem [{\citenamefont {Mezger}\ \emph {et~al.}(2008)\citenamefont {Mezger},
  \citenamefont {Schr{\"o}der}, \citenamefont {Reichert}, \citenamefont
  {Schramm}, \citenamefont {Okasinski}, \citenamefont {Sch{\"o}der},
  \citenamefont {Honkim{\"a}ki}, \citenamefont {Deutsch}, \citenamefont {Ocko},
  \citenamefont {Ralston}, \citenamefont {Rohwerder}, \citenamefont
  {Stratmann},\ and\ \citenamefont {Dosch}}]{dosch08}%
  \BibitemOpen
  \bibfield  {author} {\bibinfo {author} {\bibfnamefont {Markus}\ \bibnamefont
  {Mezger}}, \bibinfo {author} {\bibfnamefont {Heiko}\ \bibnamefont
  {Schr{\"o}der}}, \bibinfo {author} {\bibfnamefont {Harald}\ \bibnamefont
  {Reichert}}, \bibinfo {author} {\bibfnamefont {Sebastian}\ \bibnamefont
  {Schramm}}, \bibinfo {author} {\bibfnamefont {John~S.}\ \bibnamefont
  {Okasinski}}, \bibinfo {author} {\bibfnamefont {Sebastian}\ \bibnamefont
  {Sch{\"o}der}}, \bibinfo {author} {\bibfnamefont {Veijo}\ \bibnamefont
  {Honkim{\"a}ki}}, \bibinfo {author} {\bibfnamefont {Moshe}\ \bibnamefont
  {Deutsch}}, \bibinfo {author} {\bibfnamefont {Benjamin~M.}\ \bibnamefont
  {Ocko}}, \bibinfo {author} {\bibfnamefont {John}\ \bibnamefont {Ralston}},
  \bibinfo {author} {\bibfnamefont {Michael}\ \bibnamefont {Rohwerder}},
  \bibinfo {author} {\bibfnamefont {Martin}\ \bibnamefont {Stratmann}}, \ and\
  \bibinfo {author} {\bibfnamefont {Helmut}\ \bibnamefont {Dosch}},\ }\bibfield
   {title} {\enquote {\bibinfo {title} {Molecular layering of fluorinated ionic
  liquids at a charged sapphire (0001) surface},}\ }\href {\doibase
  10.1126/science.1164502} {\bibfield  {journal} {\bibinfo  {journal}
  {Science}\ }\textbf {\bibinfo {volume} {322}},\ \bibinfo {pages} {424--428}
  (\bibinfo {year} {2008})}\BibitemShut {NoStop}%
\bibitem [{\citenamefont {Atkin}\ and\ \citenamefont {Warr}(2007)}]{atkin07}%
  \BibitemOpen
  \bibfield  {author} {\bibinfo {author} {\bibfnamefont {Rob}\ \bibnamefont
  {Atkin}}\ and\ \bibinfo {author} {\bibfnamefont {Gregory~G.}\ \bibnamefont
  {Warr}},\ }\bibfield  {title} {\enquote {\bibinfo {title} {Structure in
  confined room-temperature ionic liquids},}\ }\href {\doibase
  10.1021/jp067420g} {\bibfield  {journal} {\bibinfo  {journal} {The Journal of
  Physical Chemistry C}\ }\textbf {\bibinfo {volume} {111}},\ \bibinfo {pages}
  {5162--5168} (\bibinfo {year} {2007})}\BibitemShut {NoStop}%
\bibitem [{\citenamefont {Perkin}\ \emph {et~al.}(2010)\citenamefont {Perkin},
  \citenamefont {Albrecht},\ and\ \citenamefont {Klein}}]{perkin10}%
  \BibitemOpen
  \bibfield  {author} {\bibinfo {author} {\bibfnamefont {Susan}\ \bibnamefont
  {Perkin}}, \bibinfo {author} {\bibfnamefont {Tim}\ \bibnamefont {Albrecht}},
  \ and\ \bibinfo {author} {\bibfnamefont {Jacob}\ \bibnamefont {Klein}},\
  }\bibfield  {title} {\enquote {\bibinfo {title} {Layering and shear
  properties of an ionic liquid{,} 1-ethyl-3-methylimidazolium ethylsulfate{,}
  confined to nano-films between mica surfaces},}\ }\href {\doibase
  10.1039/B920571C} {\bibfield  {journal} {\bibinfo  {journal} {Phys. Chem.
  Chem. Phys.}\ }\textbf {\bibinfo {volume} {12}},\ \bibinfo {pages}
  {1243--1247} (\bibinfo {year} {2010})}\BibitemShut {NoStop}%
\bibitem [{\citenamefont {Zhou}\ \emph {et~al.}(2012)\citenamefont {Zhou},
  \citenamefont {Rouha}, \citenamefont {Feng}, \citenamefont {Lee},
  \citenamefont {Docherty}, \citenamefont {Fenter}, \citenamefont {Cummings},
  \citenamefont {Fulvio}, \citenamefont {Dai}, \citenamefont {McDonough},
  \citenamefont {Presser},\ and\ \citenamefont {Gogotsi}}]{zhou12}%
  \BibitemOpen
  \bibfield  {author} {\bibinfo {author} {\bibfnamefont {Hua}\ \bibnamefont
  {Zhou}}, \bibinfo {author} {\bibfnamefont {Michael}\ \bibnamefont {Rouha}},
  \bibinfo {author} {\bibfnamefont {Guang}\ \bibnamefont {Feng}}, \bibinfo
  {author} {\bibfnamefont {Sang~Soo}\ \bibnamefont {Lee}}, \bibinfo {author}
  {\bibfnamefont {Hugh}\ \bibnamefont {Docherty}}, \bibinfo {author}
  {\bibfnamefont {Paul}\ \bibnamefont {Fenter}}, \bibinfo {author}
  {\bibfnamefont {Peter~T.}\ \bibnamefont {Cummings}}, \bibinfo {author}
  {\bibfnamefont {Pasquale~F.}\ \bibnamefont {Fulvio}}, \bibinfo {author}
  {\bibfnamefont {Sheng}\ \bibnamefont {Dai}}, \bibinfo {author} {\bibfnamefont
  {John}\ \bibnamefont {McDonough}}, \bibinfo {author} {\bibfnamefont {Volker}\
  \bibnamefont {Presser}}, \ and\ \bibinfo {author} {\bibfnamefont {Yury}\
  \bibnamefont {Gogotsi}},\ }\bibfield  {title} {\enquote {\bibinfo {title}
  {Nanoscale perturbations of room temperature ionic liquid structure at
  charged and uncharged interfaces},}\ }\href {\doibase 10.1021/nn303355b}
  {\bibfield  {journal} {\bibinfo  {journal} {ACS Nano}\ }\textbf {\bibinfo
  {volume} {6}},\ \bibinfo {pages} {9818--9827} (\bibinfo {year} {2012})},\
  \bibinfo {note} {pMID: 23092400}\BibitemShut {NoStop}%
\bibitem [{\citenamefont {Smith}\ \emph {et~al.}(2013)\citenamefont {Smith},
  \citenamefont {Lovelock}, \citenamefont {Gosvami}, \citenamefont {Licence},
  \citenamefont {Dolan}, \citenamefont {Welton},\ and\ \citenamefont
  {Perkin}}]{smith13}%
  \BibitemOpen
  \bibfield  {author} {\bibinfo {author} {\bibfnamefont {Alexander~M.}\
  \bibnamefont {Smith}}, \bibinfo {author} {\bibfnamefont {Kevin R.~J.}\
  \bibnamefont {Lovelock}}, \bibinfo {author} {\bibfnamefont {Nitya~Nand}\
  \bibnamefont {Gosvami}}, \bibinfo {author} {\bibfnamefont {Peter}\
  \bibnamefont {Licence}}, \bibinfo {author} {\bibfnamefont {Andrew}\
  \bibnamefont {Dolan}}, \bibinfo {author} {\bibfnamefont {Tom}\ \bibnamefont
  {Welton}}, \ and\ \bibinfo {author} {\bibfnamefont {Susan}\ \bibnamefont
  {Perkin}},\ }\bibfield  {title} {\enquote {\bibinfo {title} {Monolayer to
  bilayer structural transition in confined pyrrolidinium-based ionic
  liquids},}\ }\href {\doibase 10.1021/jz301965d} {\bibfield  {journal}
  {\bibinfo  {journal} {The Journal of Physical Chemistry Letters}\ }\textbf
  {\bibinfo {volume} {4}},\ \bibinfo {pages} {378--382} (\bibinfo {year}
  {2013})},\ \bibinfo {note} {pMID: 26281727}\BibitemShut {NoStop}%
\bibitem [{\citenamefont {{Cheng}}\ \emph {et~al.}(2016)\citenamefont
  {{Cheng}}, \citenamefont {{Dienemann}}, \citenamefont {{Stock}},
  \citenamefont {{Merola}}, \citenamefont {{Chen}},\ and\ \citenamefont
  {{Valtiner}}}]{cheng16}%
  \BibitemOpen
  \bibfield  {author} {\bibinfo {author} {\bibfnamefont {H.-W.}\ \bibnamefont
  {{Cheng}}}, \bibinfo {author} {\bibfnamefont {J.-N.}\ \bibnamefont
  {{Dienemann}}}, \bibinfo {author} {\bibfnamefont {P.}~\bibnamefont
  {{Stock}}}, \bibinfo {author} {\bibfnamefont {C.}~\bibnamefont {{Merola}}},
  \bibinfo {author} {\bibfnamefont {Y.-J.}\ \bibnamefont {{Chen}}}, \ and\
  \bibinfo {author} {\bibfnamefont {M.}~\bibnamefont {{Valtiner}}},\ }\bibfield
   {title} {\enquote {\bibinfo {title} {{The Effect of Water and Confinement on
  Self-Assembly of Imidazolium Based Ionic Liquids at Mica Interfaces}},}\
  }\href {\doibase 10.1038/srep30058} {\bibfield  {journal} {\bibinfo
  {journal} {Scientific Reports}\ }\textbf {\bibinfo {volume} {6}},\ \bibinfo
  {eid} {30058} (\bibinfo {year} {2016})}\BibitemShut {NoStop}%
\bibitem [{\citenamefont {Smith}\ \emph {et~al.}(2017)\citenamefont {Smith},
  \citenamefont {Lee},\ and\ \citenamefont {Perkin}}]{smith17}%
  \BibitemOpen
  \bibfield  {author} {\bibinfo {author} {\bibfnamefont {Alexander~M.}\
  \bibnamefont {Smith}}, \bibinfo {author} {\bibfnamefont {Alpha~A.}\
  \bibnamefont {Lee}}, \ and\ \bibinfo {author} {\bibfnamefont {Susan}\
  \bibnamefont {Perkin}},\ }\bibfield  {title} {\enquote {\bibinfo {title}
  {Switching the structural force in ionic liquid-solvent mixtures by varying
  composition},}\ }\href {\doibase 10.1103/PhysRevLett.118.096002} {\bibfield
  {journal} {\bibinfo  {journal} {Phys. Rev. Lett.}\ }\textbf {\bibinfo
  {volume} {118}},\ \bibinfo {pages} {096002} (\bibinfo {year}
  {2017})}\BibitemShut {NoStop}%
\bibitem [{\citenamefont {Varela}\ \emph {et~al.}(2015)\citenamefont {Varela},
  \citenamefont {Méndez-Morales}, \citenamefont {Carrete}, \citenamefont
  {Gómez-González}, \citenamefont {Docampo-Álvarez}, \citenamefont
  {Gallego}, \citenamefont {Cabeza},\ and\ \citenamefont {Russina}}]{varela15}%
  \BibitemOpen
  \bibfield  {author} {\bibinfo {author} {\bibfnamefont {L.M.}\ \bibnamefont
  {Varela}}, \bibinfo {author} {\bibfnamefont {T.}~\bibnamefont
  {Méndez-Morales}}, \bibinfo {author} {\bibfnamefont {J.}~\bibnamefont
  {Carrete}}, \bibinfo {author} {\bibfnamefont {V.}~\bibnamefont
  {Gómez-González}}, \bibinfo {author} {\bibfnamefont {B.}~\bibnamefont
  {Docampo-Álvarez}}, \bibinfo {author} {\bibfnamefont {L.J.}\ \bibnamefont
  {Gallego}}, \bibinfo {author} {\bibfnamefont {O.}~\bibnamefont {Cabeza}}, \
  and\ \bibinfo {author} {\bibfnamefont {O.}~\bibnamefont {Russina}},\
  }\bibfield  {title} {\enquote {\bibinfo {title} {Solvation of molecular
  cosolvents and inorganic salts in ionic liquids: A review of molecular
  dynamics simulations},}\ }\href {\doibase
  http://doi.org/10.1016/j.molliq.2015.06.036} {\bibfield  {journal} {\bibinfo
  {journal} {Journal of Molecular Liquids}\ }\textbf {\bibinfo {volume} {210,
  Part B}},\ \bibinfo {pages} {178 -- 188} (\bibinfo {year} {2015})},\ \bibinfo
  {note} {mesoscopic structure and dynamics in ionic liquids}\BibitemShut
  {NoStop}%
\bibitem [{\citenamefont {Gavish}\ and\ \citenamefont
  {Yochelis}(2016)}]{Gavish16}%
  \BibitemOpen
  \bibfield  {author} {\bibinfo {author} {\bibfnamefont {Nir}\ \bibnamefont
  {Gavish}}\ and\ \bibinfo {author} {\bibfnamefont {Arik}\ \bibnamefont
  {Yochelis}},\ }\bibfield  {title} {\enquote {\bibinfo {title} {Theory of
  phase separation and polarization for pure ionic liquids},}\ }\href {\doibase
  10.1021/acs.jpclett.6b00370} {\bibfield  {journal} {\bibinfo  {journal} {J.
  Phys. Chem. Lett.}\ }\textbf {\bibinfo {volume} {7}},\ \bibinfo {pages}
  {1121−1126} (\bibinfo {year} {2016})}\BibitemShut {NoStop}%
\bibitem [{\citenamefont {Démery}\ \emph {et~al.}(2012)\citenamefont
  {Démery}, \citenamefont {Dean}, \citenamefont {Hammant}, \citenamefont
  {Horgan},\ and\ \citenamefont {Podgornik}}]{Demery12a}%
  \BibitemOpen
  \bibfield  {author} {\bibinfo {author} {\bibfnamefont {Vincent}\ \bibnamefont
  {Démery}}, \bibinfo {author} {\bibfnamefont {David~S.}\ \bibnamefont
  {Dean}}, \bibinfo {author} {\bibfnamefont {Thomas~C.}\ \bibnamefont
  {Hammant}}, \bibinfo {author} {\bibfnamefont {Ronald~R.}\ \bibnamefont
  {Horgan}}, \ and\ \bibinfo {author} {\bibfnamefont {Rudolf}\ \bibnamefont
  {Podgornik}},\ }\bibfield  {title} {\enquote {\bibinfo {title} {Overscreening
  in 1d lattice coulomb gas model of ionic liquids},}\ }\href@noop {}
  {\bibfield  {journal} {\bibinfo  {journal} {Europhys. Lett.}\ }\textbf
  {\bibinfo {volume} {97}},\ \bibinfo {pages} {28004} (\bibinfo {year}
  {2012})}\BibitemShut {NoStop}%
\bibitem [{\citenamefont {Bazant}\ \emph {et~al.}(2011)\citenamefont {Bazant},
  \citenamefont {Storey},\ and\ \citenamefont {Kornyshev}}]{bazant11}%
  \BibitemOpen
  \bibfield  {author} {\bibinfo {author} {\bibfnamefont {Martin~Z.}\
  \bibnamefont {Bazant}}, \bibinfo {author} {\bibfnamefont {Brian~D.}\
  \bibnamefont {Storey}}, \ and\ \bibinfo {author} {\bibfnamefont {Alexei~A.}\
  \bibnamefont {Kornyshev}},\ }\bibfield  {title} {\enquote {\bibinfo {title}
  {Double layer in ionic liquids: Overscreening versus crowding},}\ }\href
  {\doibase 10.1103/PhysRevLett.106.046102} {\bibfield  {journal} {\bibinfo
  {journal} {Phys. Rev. Lett.}\ }\textbf {\bibinfo {volume} {106}},\ \bibinfo
  {pages} {046102} (\bibinfo {year} {2011})}\BibitemShut {NoStop}%
\bibitem [{\citenamefont {Naji}\ \emph {et~al.}(2013)\citenamefont {Naji},
  \citenamefont {Kandu\v{c}}, \citenamefont {Forsman},\ and\ \citenamefont
  {Podgornik}}]{Perspective}%
  \BibitemOpen
  \bibfield  {author} {\bibinfo {author} {\bibfnamefont {A.}~\bibnamefont
  {Naji}}, \bibinfo {author} {\bibfnamefont {M.}~\bibnamefont {Kandu\v{c}}},
  \bibinfo {author} {\bibfnamefont {J.}~\bibnamefont {Forsman}}, \ and\
  \bibinfo {author} {\bibfnamefont {R.}~\bibnamefont {Podgornik}},\ }\bibfield
  {title} {\enquote {\bibinfo {title} {Perspective: Coulomb fluids -- weak
  coupling, strong coupling, in between and beyond},}\ }\href@noop {}
  {\bibfield  {journal} {\bibinfo  {journal} {J. Chem. Phys.}\ }\textbf
  {\bibinfo {volume} {139}},\ \bibinfo {pages} {150901} (\bibinfo {year}
  {2013})}\BibitemShut {NoStop}%
\bibitem [{\citenamefont {Santangelo}(2006)}]{santangelo06}%
  \BibitemOpen
  \bibfield  {author} {\bibinfo {author} {\bibfnamefont {Christian~D.}\
  \bibnamefont {Santangelo}},\ }\bibfield  {title} {\enquote {\bibinfo {title}
  {Computing counterion densities at intermediate coupling},}\ }\href {\doibase
  10.1103/PhysRevE.73.041512} {\bibfield  {journal} {\bibinfo  {journal} {Phys.
  Rev. E}\ }\textbf {\bibinfo {volume} {73}},\ \bibinfo {pages} {041512}
  (\bibinfo {year} {2006})}\BibitemShut {NoStop}%
\bibitem [{\citenamefont {Paillusson}\ and\ \citenamefont
  {Blossey}(2010)}]{paillusson10}%
  \BibitemOpen
  \bibfield  {author} {\bibinfo {author} {\bibfnamefont {Fabien}\ \bibnamefont
  {Paillusson}}\ and\ \bibinfo {author} {\bibfnamefont {Ralf}\ \bibnamefont
  {Blossey}},\ }\bibfield  {title} {\enquote {\bibinfo {title} {Slits, plates,
  and poisson-boltzmann theory in a local formulation of nonlocal
  electrostatics},}\ }\href {\doibase 10.1103/PhysRevE.82.052501} {\bibfield
  {journal} {\bibinfo  {journal} {Phys. Rev. E}\ }\textbf {\bibinfo {volume}
  {82}},\ \bibinfo {pages} {052501} (\bibinfo {year} {2010})}\BibitemShut
  {NoStop}%
\bibitem [{\citenamefont {Maggs}\ and\ \citenamefont
  {Podgornik}(2016)}]{maggs16}%
  \BibitemOpen
  \bibfield  {author} {\bibinfo {author} {\bibfnamefont {A.~C.}\ \bibnamefont
  {Maggs}}\ and\ \bibinfo {author} {\bibfnamefont {R.}~\bibnamefont
  {Podgornik}},\ }\bibfield  {title} {\enquote {\bibinfo {title} {General
  theory of asymmetric steric interactions in electrostatic double layers},}\
  }\href {\doibase 10.1039/C5SM01757B} {\bibfield  {journal} {\bibinfo
  {journal} {Soft Matter}\ }\textbf {\bibinfo {volume} {12}},\ \bibinfo {pages}
  {1219--1229} (\bibinfo {year} {2016})}\BibitemShut {NoStop}%
\bibitem [{\citenamefont {Borukhov}\ \emph {et~al.}(1997)\citenamefont
  {Borukhov}, \citenamefont {Andelman},\ and\ \citenamefont
  {Orland}}]{borukhov97}%
  \BibitemOpen
  \bibfield  {author} {\bibinfo {author} {\bibfnamefont {Itamar}\ \bibnamefont
  {Borukhov}}, \bibinfo {author} {\bibfnamefont {David}\ \bibnamefont
  {Andelman}}, \ and\ \bibinfo {author} {\bibfnamefont {Henri}\ \bibnamefont
  {Orland}},\ }\bibfield  {title} {\enquote {\bibinfo {title} {Steric effects
  in electrolytes: A modified poisson-boltzmann equation},}\ }\href {\doibase
  10.1103/PhysRevLett.79.435} {\bibfield  {journal} {\bibinfo  {journal} {Phys.
  Rev. Lett.}\ }\textbf {\bibinfo {volume} {79}},\ \bibinfo {pages} {435--438}
  (\bibinfo {year} {1997})}\BibitemShut {NoStop}%
\bibitem [{\citenamefont {Borukhov}\ \emph {et~al.}(2000)\citenamefont
  {Borukhov}, \citenamefont {Andelman},\ and\ \citenamefont
  {Orland}}]{borukhov00}%
  \BibitemOpen
  \bibfield  {author} {\bibinfo {author} {\bibfnamefont {I.}~\bibnamefont
  {Borukhov}}, \bibinfo {author} {\bibfnamefont {D.}~\bibnamefont {Andelman}},
  \ and\ \bibinfo {author} {\bibfnamefont {H.}~\bibnamefont {Orland}},\
  }\bibfield  {title} {\enquote {\bibinfo {title} {Adsorption of large ions
  from an electrolyte solution: a modified poisson–boltzmann equation},}\
  }\href {\doibase http://doi.org/10.1016/S0013-4686(00)00576-4} {\bibfield
  {journal} {\bibinfo  {journal} {Electrochimica Acta}\ }\textbf {\bibinfo
  {volume} {46}},\ \bibinfo {pages} {221 -- 229} (\bibinfo {year}
  {2000})}\BibitemShut {NoStop}%
\bibitem [{\citenamefont {Curtin}\ and\ \citenamefont
  {Ashcroft}(1985)}]{curtin}%
  \BibitemOpen
  \bibfield  {author} {\bibinfo {author} {\bibfnamefont {W.~A.}\ \bibnamefont
  {Curtin}}\ and\ \bibinfo {author} {\bibfnamefont {N.~W.}\ \bibnamefont
  {Ashcroft}},\ }\bibfield  {title} {\enquote {\bibinfo {title}
  {Weighted-density-functional theory of inhomogeneous liquids and the freezing
  transition},}\ }\href {\doibase 10.1103/PhysRevA.32.2909} {\bibfield
  {journal} {\bibinfo  {journal} {Phys. Rev. A}\ }\textbf {\bibinfo {volume}
  {32}},\ \bibinfo {pages} {2909--2919} (\bibinfo {year} {1985})}\BibitemShut
  {NoStop}%
\bibitem [{\citenamefont {Maggs}\ and\ \citenamefont {Everaers}(2006)}]{ralfe}%
  \BibitemOpen
  \bibfield  {author} {\bibinfo {author} {\bibfnamefont {A.~C.}\ \bibnamefont
  {Maggs}}\ and\ \bibinfo {author} {\bibfnamefont {R.}~\bibnamefont
  {Everaers}},\ }\bibfield  {title} {\enquote {\bibinfo {title} {Simulating
  nanoscale dielectric response},}\ }\href {\doibase
  10.1103/PhysRevLett.96.230603} {\bibfield  {journal} {\bibinfo  {journal}
  {Phys. Rev. Lett.}\ }\textbf {\bibinfo {volume} {96}},\ \bibinfo {pages}
  {230603} (\bibinfo {year} {2006})}\BibitemShut {NoStop}%
\bibitem [{\citenamefont {{Pujos}}\ and\ \citenamefont
  {{Maggs}}(2014)}]{maggsxx}%
  \BibitemOpen
  \bibfield  {author} {\bibinfo {author} {\bibfnamefont {J.~S.}\ \bibnamefont
  {{Pujos}}}\ and\ \bibinfo {author} {\bibfnamefont {A.~C.}\ \bibnamefont
  {{Maggs}}},\ }\bibfield  {title} {\enquote {\bibinfo {title} {Legendre
  transforms for electrostatic energies},}\ }in\ \href@noop {} {\emph {\bibinfo
  {booktitle} {{Electrostatics of soft and disordered matter}}}},\ \bibinfo
  {editor} {edited by\ \bibinfo {editor} {\bibfnamefont {David~S}\ \bibnamefont
  {Dean}}, \bibinfo {editor} {\bibfnamefont {Jure}\ \bibnamefont {Dobnikar}},
  \bibinfo {editor} {\bibfnamefont {Ali}\ \bibnamefont {Naji}}, \ and\ \bibinfo
  {editor} {\bibfnamefont {Rudolf}\ \bibnamefont {Podgornik}}}\ (\bibinfo
  {publisher} {Pan Stanford},\ \bibinfo {address} {Singapore},\ \bibinfo {year}
  {2014})\BibitemShut {NoStop}%
\end{thebibliography}%

\end{document}